\documentclass[a4paper,11pt]{article}

\usepackage{amsmath}
\usepackage[margin=1.1in]{geometry}
\usepackage{graphicx}
\usepackage{color}
\usepackage[square,comma,numbers,sort&compress]{natbib}
\usepackage{caption}
\usepackage{subcaption}
\usepackage{amssymb}
\usepackage[abs]{overpic}

\usepackage{tikz}

\usepackage{my_macros}

\title{Understanding how porosity gradients can make a better filter using homogenization theory}
\author{M. P. Dalwadi$^{1,2, a}$, I. M. Griffiths$^{2, b}$, and M. Bruna$^{2, c}$ \\
\small $^{1}$ School of Life Sciences, University of Nottingham, Nottingham, NG7 2RD \\
\small $^{2}$ Mathematical Institute, University of Oxford, Oxford, OX2 6GG \\
\small $^{a}$ email: mohit.dalwadi@nottingham.ac.uk \\
\small $^{b}$ email: ian.griffiths@maths.ox.ac.uk \\
\small $^{c}$ email: bruna@maths.ox.ac.uk}

\date{23 September 2015}

\numberwithin{equation}{section}

\graphicspath{{./}{figures/}}

\begin{document}

\maketitle

\begin{abstract}
Filters whose porosity decreases with depth are often more efficient at removing solute from a fluid than filters with a uniform porosity. We investigate this phenomenon via an extension of homogenization theory that accounts for a macroscale variation in microstructure. In the first stage of the paper, we homogenize the problems of flow through a filter with a near-periodic microstructure and of solute transport due to advection, diffusion, and filter adsorption. In the second stage, we use the computationally efficient homogenized equations to investigate and quantify why porosity gradients can improve filter efficiency. We find that a porosity gradient has a much larger effect on the uniformity of adsorption than it does on the total adsorption. This allows us to understand how a decreasing porosity can lead to a greater filter efficiency, by lowering the risk of localized blocking while maintaining the rate of total contaminant removal. 
\end{abstract}

\section{Introduction}

Membrane separation is a vast industry with a wide range of applications, including water treatment \citep{zularisam2006behaviours,vandevivere1998review}, biopharmaceuticals \citep{burnouf2003nanofiltration,van2001membrane}, and food processing \citep{daufin2001recent,girard2000membrane}. For example, filters are crucial to remove waste and excess water from the blood in kidney dialysis \citep{lonsdale1982growth}, and yeast and bacteria in beer production \citep{fillaudeau2002yeast}. Despite the diverse industrial applications, an overarching goal of membrane design is to maximize the product yield. In applications where the solvent is the desired product, such as in water treatment, this is accomplished by maximizing both particle removal from the fluid suspension and filter lifespan. Mathematical modelling can offer key insight into the filtration process and operating conditions, and thus provide a cost-effective way to optimize filter design. 

A common form of membrane separation is \emph{depth filtration}, in which small particles (contaminants) are trapped within, and not just on the surface of, the porous filter material. Such filters often capture the majority of the contaminants in the initial portion of the filter while leaving the latter portions relatively unused, leading to premature clogging and reduced filtration efficiency~\citep{datta1998gradient}. In such cases, filters whose porosity decreases with depth, or \emph{porosity-graded filters}, which are an example of a functionally graded material, can improve filtration efficiency, and so are often used experimentally~\citep{burggraaf1991synthesis,barg2009processing,anderson1951filter,dickerson2005gradient,vida2012characterization}. Their increased efficiency can be qualitatively attributed to a decrease in porosity compensating for a reduction of contaminant concentration with depth, which occurs due to prior filtering. However, the mechanism behind this observation is not fully understood. Experiments are costly and moreover it is very difficult to observe the particle trapping within a filter during the filtration process directly. Instead, deductions are made only after dissecting the porous medium once filtration has ceased. For these reasons, optimizing a porosity distribution via a systematic series of experiments is impractical.

Mathematical and computational methods to model the filtration process are very useful for investigating this issue at a fraction of the full experimental cost. A summary table of Computational Fluid Dynamics (CFD) models for filtration is given in~\cite{rahimi2009cfd}. In the same paper, the authors use scanning electron microscopes to obtain a full description of a given membrane microstructure, and implement a full CFD model of particles moving within the membrane. Whilst a full CFD simulation gives excellent insight into how an individual particle is trapped, there are large computational costs associated with keeping track of all the particles within a complicated pore structure. Additionally, as in the example given above, each membrane must be scanned to accurately represent its pore structure. Although filtration occurs on the scale of the particle or pore size, it is generally the overall macroscale behaviour, such as the total mass of particles removed, which is the main concern in filter design.

An alternative approach to CFD for complex heterogeneous materials, such as filters, is to consider upscaling methods. These reduce the complexity of an equation by averaging any microscale variation while retaining the important macroscale variation. Mathematical homogenization via the method of multiple scales is an asymptotic technique often used for this upscaling. Traditionally, for this technique to be appropriate to use, there must be a periodic microstructure and the ratio between the length of the periodic cell and the length of the important macroscale variation must be small. In fluid mechanics, where the underlying flow equations can be difficult to solve, this procedure is used to determine the macroscale behaviour of fluids in complicated geometries. For example, in saturated single-phase Stokes flow past a periodic array of inert obstacles, this homogenization procedure leads to Darcy's Law~\citep{keller1980darcy}, and, in non-saturated media, homogenization techniques can be used to derive Richards' equation~\citep{Daly2015homog}.

Whilst the majority of homogenization procedures are carried out in domains whose microstructure is fully periodic, and thus would only be applicable to filters with a constant macroscale porosity, extensions to non-periodic domains have been explored. In~\citep{chernyavsky2011transport}, the steady problem of nutrient uptake past randomly placed point sinks is considered in one spatial dimension. As the governing equations can be solved if the locations of the sinks are known, significant analytic progress is made into investigating the macroscale effect of different random distributions. A comprehensive extension to near-periodic domains is developed in~\citep{richardson2011derivation}, where the authors use a general curvilinear coordinate transform to homogenize the electric potential within a beating heart, mapping the near-periodic microscale to a periodic domain. Similar approaches are used to develop homogenized equations in~\citep{penta2014effective}, where the authors consider saturated flow in a poroelastic solid with surface growth of the solid phase, which is allowed to have a macroscopic variation, and in~\citep{penta2015multiscale}, where the authors consider blood and drug exchange across capillary walls in malignant tumours.

Although these extensions to non-periodic domains are certainly significant advancements to the field, their generality means that, usually, the cell problem must be solved at every point in the macroscale, reducing the computational efficiency. This issue is bypassed in~\citep{Bruna2015diffusion}, where the multiple scales homogenization method is extended to quasi-periodic structures where the microstructure is allowed to vary slowly. In particular, the authors consider the problem of diffusion in a porous medium whose solid phase is modelled as an array of inert spherical inclusions (obstacles). The radii of these obstacles are allowed to vary spatially over the macroscale length. By imposing a specific one-parameter form on the obstacles, an explicit formulation of the macroscale equation is obtained in terms of the cell-averaged porosity, which varies in the macroscale. A notable aspect of this analysis is that a porosity variation induces a macroscale particle advection in the direction of decreasing porosity.

In this paper we employ the homogenization method developed in~\citep{richardson2011derivation,Bruna2015diffusion} to investigate porosity-graded filters. We model the filter material as a collection of spherical obstructions, whose radii vary spatially over a long lengthscale, past which a fluid with suspended contaminant particles flows due to an applied transmembrane pressure. The particles are transported via advection and diffusion, and can be trapped on the filter microstructure. In general, the accumulation of contaminant particles via trapping modifies the filter geometry, and this process eventually leads to pore blockage within the filter. Since this effect occurs within and not just on the inlet surface of the filter, the blockage is difficult to remove. Dealing with a blockage causes a significant reduction in efficiency and an increase in maintenance cost, as the filtration procedure must be halted to perform a back-flow procedure to dislodge the blockage or, more drastically, a filter replacement.

The underlying trapping or adsorption mechanism of a particle to the filter structure will, in principle, depend on the solute solubility and the filter material adsorption. Many existing models for trapping in filters use constitutive macroscale laws to determine how fluid volume throughput varies in time, which are fitted with experimental data~\citep{bowen1995steps}. In recent work by~\citep{griffiths2014combined}, the authors perform simulations of a microscale model for the blocking of a network of pores formulated from a more fundamental mechanistic level, and their results are consistent with the set of constitutive laws. However, the model in \citep{griffiths2014combined} does not account for the flow profile and is computationally expensive as it keeps track of each pore. 
Incorporating pore blockage and the subsequent filter failure is mathematically challenging to include in homogenized models, since pore growth leads to difficult free-boundary problems.

The main objective of this work is to understand and quantify how porosity-graded filters improve filter efficiency. To this end, starting from the flow and particle transport problems in the complicated domain described by a given microstructure, we employ the homogenization method developed in~\citep{richardson2011derivation,Bruna2015diffusion} to systematically determine an effective macroscale equation. Our model takes into account the effects of the microstructure on the fluid flow, as well as the transport by diffusion and advection of the contaminant particles. The trapping of particles by the filter microstructure is also included as an adsorption process. As in~\citep{Bruna2015diffusion}, we consider a solid structure composed of spherical inclusions whose radii vary slowly in the macroscale. However, now the inclusions are partially adsorbing sinks for the particle transport and inert obstacles to the flow. In order to focus on the effect of variations in porosity within a filter, in this work we assume that the contaminant particles are small and in dilute suspension within the fluid, and thus their trapping has a negligible effect on the obstacle size. This means that our model will not be able to explicitly capture pore blocking. As a result of this simplification, the flow problem decouples from the particle transport problem (but not \emph{vice versa}), and we do not have a free-boundary problem. As we consider a dilute suspension, we follow the lead of~\citep{polyakov2008depth} and impose that particle adsorption is linearly dependent on the number of particles.

This paper is divided into two stages. The first stage consists of the presentation of the full problem in \S\ref{sec: Model desc}, and the homogenization of the flow and particle transport problems to systematically obtain effective equations on the macroscale in \S\ref{sec: Homog}. This allows us to investigate the effect of a varying porosity on filtration in the second stage of the paper, where the steady-state effective equations are applied to a filter whose porosity varies in one direction only and the flow is uni-directional, in the same direction as the porosity variation. We consider a general operating regime where the filter is placed between two reservoirs. We solve the derived system numerically and analyse our results to investigate how a variation in filter porosity affects particle trapping in steady state operation. We introduce two metrics to quantify the effectiveness of a filter and, through these, we are able to determine how a porosity gradient can improve filter efficiency. Additionally, we perform an asymptotic analysis on the derived system, exploiting the small porosity gradient (a common feature in many industrial operating regimes) to determine very accurate analytic expressions at a much reduced computational cost. These tasks are all carried out in \S\ref{sec: Uni-dir problem}. We conclude in \S\ref{sec: Discussion} with an overview and discussion of the results from this paper, ending with relevant extensions based on this work.

\section{Model description}
\label{sec: Model desc}

We consider the transport of particles via advection and diffusion through a porosity-graded filter. The filter is modelled as a collection of solid obstacles, to which particles can adsorb. We describe the particle locations in terms of the concentration $\ndc(\ndx,\ndt)$, where $\ndx$ is the spatial vector coordinate and $\ndt$ is time.

The concentration field is defined within the fluid phase of the domain, where $\Regf \subset \mathbb{R}^d$ is the fluid phase and $\Rege \subset \mathbb{R}^d$ is the entire domain (which we refer to as the `porous medium'), with $d = 2$ or $3$ the number of spatial dimensions. We define the solid phase of the domain as $\Regs \subset \mathbb{R}^d$, noting that $\Regs = \Rege \setminus \Regf$. The solid phase is modelled by a collection of fixed non-overlapping $d$-dimensional balls, whose centres are located on a square or cubic lattice at a distance $\epsa \leng$ apart, where $\epsa$ is a (small) dimensionless parameter and $\leng$ is the characteristic filter length. We allow the radii of the balls to vary in space, and a ball with centre at $\ndx$ has radius $\epsa \leng \eps(\ndx)$.

A consequence of the assumption that the contaminants are small particles which are in dilute suspension is that particle--particle interactions are negligible, and the dominant interaction is the adsorption of particles on the obstacles. Thus, the particle concentration is governed by the standard advection--diffusion equation with a reactive (or partially adsorbing) Robin boundary condition, which represents a linear adsorption rate:
\begin{subequations}
\label{eq: transport problem}
\begin{alignat}{2}
\pbyp{\ndc}{\ndt} &= \ndn \bcdot \left( D \ndn \conc - \ndu \ndc \right),& \qquad  &\ndx \in \Regf , \\
- \kn \ndc &= \bs{n}\bcdot \left( D \ndn \ndc - \ndu \ndc \right),& \qquad &\ndx \in \partial \Regf.
\end{alignat}
\end{subequations}
Here $\ndn$ refers to the nabla operator with respect to $\ndx$, $D > 0$ is the constant diffusion coefficient, $\partial \Regf$ is the interface between the solid obstacles and the fluid region (hereafter known as the solid--fluid interface), $\bs{n}$ is the outward unit normal to $\partial \Regf$, $\uu(\ndx,\ndt)$ is the velocity of the fluid flow, and $\kn \geqslant 0$ is the constant particle-adsorption coefficient. There is no adsorption when $\kn = 0$, and the adsorption is instantaneous in the limit as $\kn \to \infty$.

Additionally, we assume that we have an incompressible Newtonian fluid, which satisfies Stokes equations,
\begin{subequations}
\label{eq: Flow problem}
\begin{alignat}{2}
-\ndn \ndp + \mu \ndn^2 \ndu &= \zv,& \qquad  &\ndx \in \Regf, \\
\ndn \bcdot \ndu &= 0,& \qquad  &\ndx \in \Regf, \\
\ndu &= \zv,& \qquad &\ndx \in \partial \Regf,
\end{alignat}
\end{subequations}
where $\viscosity$ is the fluid viscosity and $\ndp(\ndx,\ndt)$ is the fluid pressure.

\subsection{Dimensionless equations}
\label{sec: Dimensionless equations}

We scale the variables via $\ndx = \leng \bs{x}$, $\ndu = \velscale \uu$, $\ndt = (\leng / \velscale)t$, $\ndc = c_{\infty} \conc$, and $\ndp = (\viscosity \velscale/(\epsa^2 \leng))p$, where $\velscale$ is a characteristic velocity scale and $c_\infty$ is a characteristic concentration scale. The pressure scaling is chosen to balance the pressure gradient over the macroscale with viscous forces over the obstacle microscale. Using these scalings in~\eqref{eq: transport problem}, we obtain the dimensionless solute-transport equation
\begin{subequations}
\label{eq: transport problem ND}
\begin{alignat}{2}
\pbyp{\conc}{t} &= \nabla \bcdot \left( \Pec^{-1}\nabla \conc - \uu  \conc \right),& \qquad  &\bs{x} \in \Regf , \\
\label{eq: transport problem ND BC}
- \epsa \kk \conc &= \bs{n}\bcdot \left( \Pec^{-1}\nabla \conc - \uu  \conc \right),& \qquad &\bs{x} \in \partial \Regf,
\end{alignat}
\end{subequations}
where $\Pec = \velscale \leng / D$, the P\'{e}clet number, and $\kk = \kn/(\epsa \velscale)$ are both assumed to be $\order{1}$ parameters. As we see in \S\ref{sec: Homog}, this choice yields a distinguished limit in which all mechanisms are present at leading order in the macroscale. We note that $\kk$ may be expressed in terms of the P\'{e}clet number and the Damk\"{o}hler number $\Dam$ (which relates adsorption to diffusion) via $\kk = \Dam/(\epsa \Pec)$. The parameter $\epsa$ that premultiplies the adsorption term in~\eqref{eq: transport problem ND BC} is required to ensure a dominant balance between particle transport and adsorption, since an $\order{\epsa}$ adsorption rate over an $\order{1/\epsa}$ number of obstacles will lead to an overall $\order{1}$ adsorption rate. It is shown in~\citep{chernyavsky2011transport} that an asymptotic analysis of a similar advection--reaction--diffusion equation with $\Pec = \order{1}$ and $\kk = \order{1/\epsa}$ leads to a breakdown in the asymptotic series considered. In the limit of large $\kk$, the solute is adsorbed very quickly and, for this work, could be tracked via an initial boundary layer in time of $\order{1/\kk}$ and, if necessary, diffusive boundary layers in space of $\order{1/\sqrt{\kk}}$ (though we do not consider these sub-limits further).

\begin{figure}[t]
\centering
    	\includegraphics[width=\textwidth]{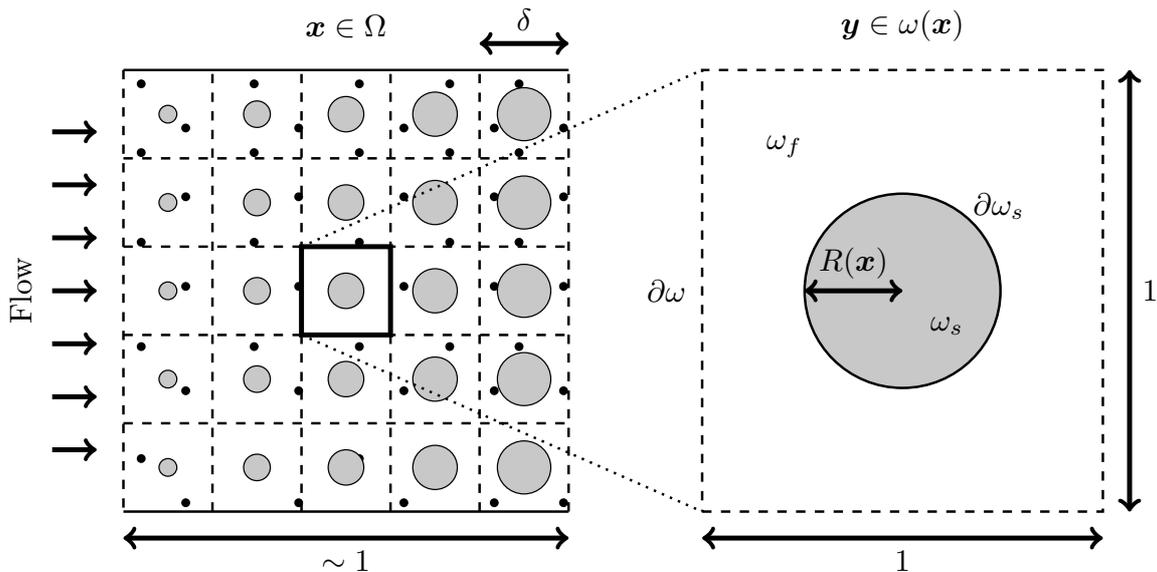}
\caption{Model schematic in two dimensions. Left: An example in which the macroscale porosity decreases in the direction of the flow. Right: A magnified view of a given cell $\rege(\bs{x})$, with microscale coordinate $\bs{y} \in \left[-\frac{1}{2},\frac{1}{2}\right]^2$.}
\label{fig: Set-up schematic}
\end{figure}

The dimensionless version of~\eqref{eq: Flow problem} is
\begin{subequations}
\label{eq: Flow problem ND}
\begin{alignat}{2}
-\nabla p + \epsa^2\nabla^2 \uu &= \zv,& \qquad  &\bs{x} \in \Regf, \\
\nabla \bcdot \uu &= 0,& \qquad  &\bs{x} \in \Regf, \\
\uu &= \zv,& \qquad &\bs{x} \in \partial \Regf.
\end{alignat}
\end{subequations}
In dimensionless units, the obstacles now form a $d$-cubic lattice of $d$-balls a distance of $\epsa$ apart, and a $d$-ball with centre at $\bs{x}$ has radius $\epsa \eps(\bs{x})$. A schematic of the dimensionless geometry is shown in the left-hand side of figure~\ref{fig: Set-up schematic}.

\section{Homogenization}
\label{sec: Homog}

In order to reduce the complexity of the problem geometry we homogenize the governing equations~\eqref{eq: transport problem ND}--\eqref{eq: Flow problem ND} using the method of multiple scales. This will allow us to obtain effective equations on a simpler macroscale domain while formally capturing the relevant information about the microscale geometry. As is standard in homogenization theory, we introduce a microscale variable $\bs{y} = \bs{x}/\epsa$ and treat $\bs{x}$ and $\bs{y}$ as independent. The extra degree of freedom this allows is removed by imposing that the solution is exactly periodic in $\bs{y}$, and hence any small variation between unit cells is captured through the macroscale variable $\bs{x}$. As shown in figure~\ref{fig: Set-up schematic}, the microscale variable $\bs y$ is defined in the unit cell $\rege(\bs{x})$, whereas the macroscale variable $\bs x$ spans the entire filter. The solid portion of the cell, occupied by the obstacle, is denoted by $\regs({\bs x})$ and the fluid portion is $\regf(\bs{x}) = \rege(\bs{x}) \backslash \regs(\bs{x})$. The boundary of the unit cell determined by the obstacle is denoted by $\partial \regs (\bs{x})$, while the outer boundary of the unit cell is $\partial \rege(\bs{x})$. Further, we consider each dependent variable as a function of both $\bs{x}$ and $\bs{y}$. Using the new variable $\bs{y}$, we transform spatial derivatives in the following manner
\begin{align}
\label{eq: spatial transform}
\nabla \mapsto \nabla_{\bs{x}} + \dfrac{1}{\epsa} \nabla_{\bs{y}},
\end{align}
where $\nabla_{\bs{x}}$ and $\nabla_{\bs{y}}$ refer to the nabla operator in the $\bs{x}$- and $\bs{y}$-coordinate systems respectively.

Our goal is to systematically reduce the geometrical complexity of the problem by deriving effective governing equations valid in the macroscale domain. For this purpose, the variables we use in our macroscale equations are quantities averaged over an entire cell $\rege(\bs{x})$. To this end, we define the porosity $\porosity(\bs{x})$ to be
\begin{align}
\label{eq: porosity}
\porosity(\bs{x}) = \frac{ |\regf(\bs{x})| }{ |\omega ({\bs x})| } \equiv |\regf(\bs{x})|,
\end{align}
since the unit cell has unit volume ($|\omega (\bs x)| = 1$), and the volumetric average concentration $\creal$ and volumetric average fluid velocity $\UU$ (also known as the Darcy velocity in porous-media formulations) as follows
\begin{subequations}
\label{eq: Volumetric average}
\begin{align}
\label{eq: Volumetric average concentration}
\creal(\bs{x},t) &= \frac{1}{|\omega (\bs{x})|} \int_{\omega (\bs{x})}  \conc(\bs{x},\bs{y},t) \, \mathrm{d}\bs{y} = \int_{\regf (\bs{x})}  \conc(\bs{x},\bs{y},t) \, \mathrm{d}\bs{y}, \\
\label{eq: Volumetric average velocity}
\UU(\bs{x},t) &=  \frac{1}{|\omega (\bs{x})|} \int_{\omega (\bs{x})}  \uu(\bs{x},\bs{y},t) \, \mathrm{d}\bs{y} =  \int_{\regf (\bs{x})}  \uu(\bs{x},\bs{y},t) \, \mathrm{d}\bs{y},
\end{align}
\end{subequations}
imposing that $\conc = \uu \equiv 0 $ in $\regs({\bs x})$. We note that, alternatively, we could have averaged over the fluid portion of the cell (using $|\regf (\bs{x})|$ in the denominator of \eqref{eq: Volumetric average}) rather than over the whole cell; this average, known as the intrinsic average, is most commonly used in the method of volume averaging \cite{whitaker:1999vj}. This would give the particle concentration in terms of the available fluid space of the membrane, rather than the overall particle concentration within a membrane. As we will see later, the volume average concentration is most convenient in our case since it allows us to directly deduce the effect of macroscopic changes in the porosity. 

We first consider the flow problem~\eqref{eq: Flow problem ND}, since it has no dependence on the concentration field, and use the ensuing results in the solute-transport problem~\eqref{eq: transport problem ND}.

\subsection{Flow problem}
\label{App: Flow problem no growth}

Under the spatial transform~\eqref{eq: spatial transform}, the flow equations~\eqref{eq: Flow problem ND} become
\begin{subequations}
\label{eq: Flow problem spatial transformed}
\begin{alignat}{2}
- \left(\epsa^{-1} \nabla_{\bs{y}} + \nabla_{\bs{x}} \right) p + \left(\nabla_{\bs{y}} + \epsa \nabla_{\bs{x}}\right)^2 \uu &= \zv,& \qquad &\bs{y} \in \regf(\bs{x}), \\
\label{eq: Flow problem spatial transformed cont}
\left(\nabla_{\bs{y}} + \epsa \nabla_{\bs{x}} \right) \bcdot \uu &= 0,& \qquad &\bs{y} \in \regf(\bs{x}), \\
\label{eq: Flow problem spatial transformed no slip}
\uu &= \zv,& \qquad &\bs{y} \in \partial \regs(\bs{x}), \\
\label{eq: Flow problem spatial transformed periodic}
\uu&, \, p \text{ periodic},& \qquad &\bs{y} \in \partial \rege(\bs{x}),
\end{alignat}
\end{subequations}
where $\partial \regs$ is the obstacle boundary, and $\partial \rege$ is the outer cell boundary as shown in figure~\ref{fig: Set-up schematic}. Expanding the flow velocity and pressure in powers of $\epsa$ as follows
\begin{align}
\label{eq: asymptotic expansions}
G(\bs{x},\bs{y},t) = G_0(\bs{x},\bs{y},t) + \epsa G_1(\bs{x},\bs{y},t) + \ldots \quad \text{as } \epsa \to 0,
\end{align}
where $G \in \{ \uu, p\}$, the flow equations~\eqref{eq: Flow problem spatial transformed} at $\order{\epsa^{-1}}$ are
\begin{align}
- \nabla_{\bs{y}}p_0 = 0.
\end{align}
Therefore, the leading-order pressure is independent of the microscale, \ie $p_0 = p_0(\bs{x},t)$, a direct consequence of our pressure scaling choice in \S\ref{sec: Dimensionless equations}.

The $\order{1}$ terms in~\eqref{eq: Flow problem spatial transformed} are given by
\begin{subequations}
\label{eq: Flow problem order 1}
\begin{alignat}{2}
- \nabla_{\bs{y}} p_1 + \nabla^2_{\bs{y}} \uu_0 &= \nabla_{\bs{x}} p_0,& \qquad &\bs{y} \in \regf(\bs{x}), \\
\label{eq: Cont eq order 1}
\nabla_{\bs{y}} \bcdot \uu_0 &= 0,& \qquad &\bs{y} \in \regf(\bs{x}), \\
\uu_0 &= \zv,& \qquad &\bs{y} \in \partial \regs(\bs{x}), \\
\uu_0&, \, p_1 \text{ periodic},& \qquad &\bs{y} \in \partial \rege(\bs{x}).
\end{alignat}
\end{subequations}
Since~\eqref{eq: Flow problem order 1} is a linear problem, its solution can be written as
\begin{subequations}
\begin{align}
\uu_0 &= - \Ku(\bs{x},\bs{y},t) \nabla_{\bs{x}} p_0, \\
p_1 &= - \pres(\bs{x},\bs{y},t) \bcdot \nabla_{\bs{x}} p_0 + \overline{p}(\bs{x},t),
\end{align}
\label{eq: u_0 and p_1 scalings}
\end{subequations}
where the matrix function $\Ku$ and the vector function $\pres$ satisfy the problem
\begin{subequations}
\label{eq: Flow problem  order 1 after scalings}
\begin{alignat}{2}
\Iu - \nabla_{\bs{y}} \pres + \nabla^2_{\bs{y}} \Ku &= \zu,& \qquad &\bs{y} \in \regf(\bs{x}), \\
\nabla_{\bs{y}} \bcdot \Ku &= \zv,& \qquad &\bs{y} \in \regf(\bs{x}), \\
\Ku &= \zu,& \qquad &\bs{y} \in \partial \regs(\bs{x}), \\
\Ku&, \, \pres \text{ periodic},& \qquad &\bs{y} \in \partial \rege(\bs{x}).
\end{alignat}
\end{subequations}
Here, $\Iu$ is the $d$-dimensional identity matrix, and the dependence of $\Ku$ and $\pres$ on $\bs{x}$ is due to the dependence of the microscale boundary $\partial \regs(\bs{x})$ on the macroscale variable.

To obtain the effective flow equation for the Darcy velocity \eqref{eq: Volumetric average velocity}, we integrate~(\ref{eq: u_0 and p_1 scalings}a) over $\regf(\bs{x})$ to deduce
\begin{subequations}
\label{eq: Darcy velocity}
\begin{align}
\label{eq: Darcy velocity order 1}
\UU({\bs x},t) & = \int_{\regf (\bs{x})}  \uu(\bs{x},\bs{y},t) \, \mathrm{d}\bs{y} \sim  -K(\porosity) \nabla_{\bs{x}} p,
\end{align}
at leading order, where $K(\porosity)$ is a scalar function defined by
\begin{align}
\label{eq: def of K}
K(\porosity)\Iu &= \int_{\regf (\bs{x})}  \Ku \, \mathrm{d}\bs{y}.
\end{align}
We note that the integral of $\Ku$ in~\eqref{eq: def of K} is a multiple of the identity matrix due to the symmetry of the cell problem described by~\eqref{eq: Flow problem  order 1 after scalings}. This would no longer be the case if, instead of a spherical obstacle, we had considered an anisotropic obstacle. Integrating~\eqref{eq: Flow problem spatial transformed cont} over $\regf$, and using the transport theorem (outlined in Appendix~\ref{App: TT}) in conjunction with the periodic and no-slip boundary conditions~\eqref{eq: Flow problem spatial transformed no slip}--\eqref{eq: Flow problem spatial transformed periodic}, we obtain the following incompressibility condition for the macroscale flow velocity
\begin{align}
\label{eq: Incomp}
\nabla_{\bs{x}} \bcdot \UU = 0.
\end{align}
\end{subequations}
Therefore, the homogenization procedure of the Stokes velocity in the cell problem~\eqref{eq: Flow problem ND} via multiple scales yields Darcy's law~\eqref{eq: Darcy velocity} for the macroscopic velocity $\UU$ as expected \cite{Cushman:2002fk}.

\subsection{Solute-transport problem}

We now perform a similar homogenization for the solute-transport problem. While the Dirichlet boundary conditions for the flow problem (\ref{eq: Flow problem ND}c) are invariant to the multiple-scales transformation, this is not true for the transformation of the Robin boundary condition (\ref{eq: transport problem ND}b) in the solute-transport problem. Following~\citep{Bruna2015diffusion,richardson2011derivation}, we address this issue by introducing the relation
\begin{align}
\label{eq: chi}
\chi(\bs{x},\bs{y}) = \eps(\bs{x}) - \|\bs{y}\|,
\end{align}
where $\chi(\bs{x},\bs{y}) = 0$ defines the solid--fluid interface.\footnote{Writing $\eps(\bs{x})$ in \eqref{eq: chi} assumes that $R$ is a continuous function of the macroscale $\bs{x}$, rather than a piecewise constant function evaluated at the centre of the relevant unit cell. This simplifies the subsequent analysis while affecting only the boundary condition \eqref{eq: transport problem spatial transform BC} at higher orders than we need to consider. As a result, our final leading-order macroscale equation \eqref{eq: effective volumetric equation} is unchanged by employing this simplification.} Using the gradient transform~\eqref{eq: spatial transform}, the normal vector $\bs{n}$ in~\eqref{eq: transport problem ND BC} becomes
\begin{align}
\label{eq: normal in microscale}
\bs{n} = \dfrac{\bs{n}_{\bs{y}} + \epsa \nabla_{\bs{x}} \eps}{\|\bs{n}_{\bs{y}} + \epsa \nabla_{\bs{x}} \eps\|},
\end{align}
where $\bs{n}_{\bs{y}} = -\bs{y} / \|\bs{y}\|$ is the outward unit normal on the obstacle boundary $\partial \regs(\bs{x})$, and $\epsa \nabla_{\bs{x}} \eps$ accounts for the macroscale effect of varying obstacle size (see \cite{richardson2011derivation} for details).

Under the spatial transforms~\eqref{eq: spatial transform} and~\eqref{eq: normal in microscale}, the solute-transport problem~\eqref{eq: transport problem ND} becomes
\begin{subequations}
\label{eq: transport problem spatial transform}
\begin{alignat}{2}
\label{eq: transport problem spatial transform eq}
\epsa^2 \pbyp{\conc}{t} &= \left(\nabla_{\bs{y}} + \epsa \nabla_{\bs{x}}\right) \bcdot \left(\Pec^{-1} \left(\nabla_{\bs{y}} + \epsa \nabla_{\bs{x}} \right) \conc - \epsa \uu \conc \right),& \qquad &\bs{y} \in \regf(\bs{x}), \\
\label{eq: transport problem spatial transform BC}
- \epsa^2 \kk \conc &= \left(\bs{n}_{\bs{y}} + \epsa \nabla_{\bs{x}} \eps \right) \bcdot\left( \Pec^{-1} \left(\nabla_{\bs{y}} + \epsa \nabla_{\bs{x}}\right) \conc  - \epsa \uu \conc \right) + \order{\epsa^3},& \qquad &\bs{y} \in \partial \regs(\bs{x}), \\
\conc &\text{ periodic},& \qquad &\bs{y} \in \partial \rege(\bs{x}).
\end{alignat}
\end{subequations}
Expanding $\conc(\bs{x},\bs{y},t)$ in powers of $\epsa$ as in~\eqref{eq: asymptotic expansions} and equating powers of $\epsa$, the $\order{1}$ terms are
\begin{subequations}
\label{eq: transport problem order 1}
\begin{alignat}{5}
&0 = \nabla_{\bs{y}}^2 \conc_0, & \qquad &\bs{y} \in \regf(\bs{x}), \\
&0 = \bs{n}_{\bs{y}} \bcdot \nabla_{\bs{y}} \conc_0,& \qquad &\bs{y} \in \partial \regs(\bs{x}), \\
&\conc_0  \text{ periodic},& \qquad &\bs{y} \in \partial \rege(\bs{x}).
\end{alignat}
\end{subequations}
Therefore, we deduce that $\conc_0$ is independent of the microscale, \ie $\conc_0 = \conc_0(\bs{x},t)$. The $\order{\epsa}$ terms in~\eqref{eq: transport problem spatial transform} are
\begin{subequations}
\label{eq: transport problem order delta}
\begin{alignat}{5}
0 &= \nabla_{\bs{y}}^2 \conc_1, & \qquad &\bs{y} \in \regf(\bs{x}), \\
-\bs{n}_{\bs{y}} \bcdot \nabla_{\bs{x}} \conc_0 &= \bs{n}_{\bs{y}} \bcdot \nabla_{\bs{y}} \conc_1, & \qquad &\bs{y} \in \partial \regs(\bs{x}), \\
\conc_1 &\text{ periodic},& \qquad &\bs{y} \in \partial \rege(\bs{x}),
\end{alignat}
\end{subequations}
using the continuity equation for the flow \eqref{eq: Cont eq order 1}. The system~\eqref{eq: transport problem order delta} is linear in $\conc_1$, which allows us to write $\conc_1$ as
\begin{align}
\label{eq: conc_1 in terms of gamma}
\conc_1(\bs{x},\bs{y},t) = - \gam(\bs{x},\bs{y}) \bcdot \nabla_{\bs{x}} \conc_0(\bs{x},t),
\end{align}
where the components of the function $\gam$ satisfy
\begin{subequations}
\label{eq: cell transport problem}
\begin{alignat}{5}
0 &= \nabla_{\bs{y}}^2 \Gamma_i, & \qquad &\bs{y} \in \regf(\bs{x}),\\
n_{y,i} &= \bs{n}_{\bs{y}} \bcdot \nabla_{\bs{y}} \Gamma_i, & \qquad &\bs{y} \in \partial \regs(\bs{x}), \\
\Gamma_i & \text{ periodic},& \qquad &\bs{y} \in \partial \rege(\bs{x}),
\end{alignat}
\end{subequations}
and $n_{y,i}$ is the $i{\text{th}}$ component of the unit vector $\bs{n}_{\bs{y}}$.

The $\order{\epsa^2}$ terms in~\eqref{eq: transport problem spatial transform} are
\begin{subequations}
\label{eq: transport problem order delta^2}
\begin{alignat}{5}
\label{eq: transport problem order delta^2 P}
\pbyp{\conc_0}{t} &= \nabla_{\bs{y}} \bcdot \left( \Pec^{-1} \left( \nabla_{\bs{y}} \conc_2  + \nabla_{\bs{x}} \conc_1 \right) - \uu_1 \conc_0 - \uu_0 \conc_1 \right) \notag \\
&\quad + \nabla_{\bs{x}} \bcdot \left(\Pec^{-1}\left( \nabla_{\bs{y}} \conc_1  + \nabla_{\bs{x}} \conc_0 \right) - \uu_0 \conc_0 \right), & \qquad &\bs{y} \in \regf(\bs{x}), \\
\label{eq: transport problem order delta^2 BC}
- \kk \conc_0 &= \bs{n}_{\bs{y}} \bcdot \left( \Pec^{-1} \left( \nabla_{\bs{y}} \conc_2  + \nabla_{\bs{x}} \conc_1 \right) - \uu_1 \conc_0 - \uu_0 \conc_1 \right) \notag \\
&\quad + \nabla_{\bs{x}} \eps \bcdot \left( \Pec^{-1} \left(\nabla_{\bs{y}} \conc_1  + \nabla_{\bs{x}} \conc_0 \right) - \uu_0 \conc_0 \right), & \qquad &\bs{y} \in \partial \regs(\bs{x}), \\
\label{eq: transport problem order delta^2 periodic}
\conc_2 & \text{ periodic},& \qquad &\bs{y} \in \partial \rege(\bs{x}).
\end{alignat}
\end{subequations}
Integrating~\eqref{eq: transport problem order delta^2 P} over $\regf$ and applying the boundary conditions~\eqref{eq: transport problem order delta^2 BC}--\eqref{eq: transport problem order delta^2 periodic} gives
\begin{align}
\label{eq: sec cond in terms of integrals pre-RTT}
\begin{aligned}
\int_{\regf(\bs{x})} \! \pbyp{\conc_0}{t} \, \mathrm{d}\bs{y} = & \int_{\regf(\bs{x})} \! \nabla_{\bs{x}} \bcdot \left(\Pec^{-1} \left(\nabla_{\bs{y}} \conc_1  + \nabla_{\bs{x}} \conc_0 \right)  - \uu_0 \conc_0 \right)\,\mathrm{d}\bs{y}\\
& -\int_{\partial \regs(\bs{x})} \! \nabla_{\bs{x}} \eps \bcdot \left(\Pec^{-1} \left(\nabla_{\bs{y}} \conc_1  + \nabla_{\bs{x}} \conc_0 \right) - \uu_0 \conc_0 \right) \, \mathrm{d}s - \int_{\partial \regs(\bs{x})} \! \kk \conc_0 \, \mathrm{d}s,
\end{aligned}
\end{align}
where $\mathrm{d}s$ denotes the differential element of the obstacle surface $\partial \regs(\bs{x})$. The first two terms on the right-hand side of~\eqref{eq: sec cond in terms of integrals pre-RTT} can be simplified using the transport theorem as stated in \eqref{eq: TT}, giving
\begin{align}
\label{eq: sec cond in terms of integrals}
\int_{\regf(\bs{x})} \! \pbyp{\conc_0}{t} \, \mathrm{d}\bs{y} = \nabla_{\bs{x}} \bcdot \int_{\regf(\bs{x})} \! \left( \Pec^{-1}  \left(\nabla_{\bs{y}} \conc_1  + \nabla_{\bs{x}} \conc_0 \right) - \uu_0 \conc_0 \right) \, \mathrm{d}\bs{y} - \int_{\partial \regs(\bs{x})} \! \kk \conc_0 \, \mathrm{d}s.
\end{align}
Using that $c_0= c_0({\bs x},t)$, $|{\regf}({\bs x})| = \phi({\bs x})$ and \eqref{eq: conc_1 in terms of gamma}, \eqref{eq: sec cond in terms of integrals} reduces to
\begin{align}
\label{eq: effective intrinsic equation}
\phi \pbyp{\conc_0}{t} = \nabla_{\bs{x}} \bcdot \left( \Pec^{-1} \int_{\regf(\porosity)} ( \Iu - \! \Jt) \, \mathrm{d}\bs{y}  \, \nabla_{\bs{x}} \conc_0 
- \int_{\regf(\porosity)}  \uu_0 \, \mathrm{d}\bs{y} \, \conc_0 \right) - |\partial \regs(\porosity)| \kk \conc_0,
\end{align}
where $(\Jt)_{ij} = \partial \Gamma_j / \partial y_i$ is the transpose of the Jacobian matrix of $\gam$, the solution to the cell problem \eqref{eq: cell transport problem}. From now on, we simply write $\porosity$ but it should be understood that the porosity will be, in general, a function of $\bs x$. In a similar manner, we use $\regf(\porosity)$ instead of $\regf(\bs{x})$ henceforth.

To express \eqref{eq: effective intrinsic equation} in terms of the volumetric average concentration $\creal ({\bs x},t)$ defined in \eqref{eq: Volumetric average concentration}, we note that $\creal({\bs x},t) \sim \int_{\regf(\porosity)} \conc_0 \mathrm{d}\bs{y} = |\regf(\bs{x})| \conc_0({\bs x},t)$ at leading order in $\epsa$. Using this relation and \eqref{eq: Darcy velocity order 1}, we find that
\begin{subequations}
\label{eq: effective final problem}
\begin{align}
\label{eq: effective volumetric equation}
\pbyp{\creal}{t} = \nabla_{\bs{x}} \bcdot \left(\Deff(\porosity) \nabla_{\bs{x}} \creal - \dfrac{\creal}{\porosity}\left(\UU(\porosity) + \Deff(\porosity) \nabla_{\bs{x}} \porosity \right)\right) - f(\porosity) \creal,
\end{align}
at leading order in $\epsa$, where the effective diffusion coefficient is
\begin{align}
\label{eq: effective coefficient matrices}
\Deff(\porosity) \Iu = \Pec^{-1} \left( \Iu - \dfrac{1}{\porosity}\int_{\regf(\porosity)} \! \Jt \, \mathrm{d}\bs{y} \right),
\end{align}
and the effective adsorption coefficient is 
\begin{align}
\label{eq: particular sink function}
f(\porosity) = \kk \dfrac{|\partial \regs(\porosity)|}{\porosity} = \kk\dfrac{ d (1 - \porosity)}{\porosity}\left(\dfrac{V_d}{1- \porosity}\right)^{1/d},
\end{align}
\end{subequations}
using the fact that $\porosity(\bs{x}) = 1 - V_d \eps(\bs{x})^d$ for spherical obstacles, where $V_d$ is the volume of the unit ball with dimension $d$, so $V_2 = \pi$ and $V_3 = 4 \pi / 3$, and $|\partial \regs(\porosity)| = d V_d \eps(\bs{x})^{d-1}$.
We recall that the Darcy velocity that appears in \eqref{eq: effective volumetric equation} is given by $\UU({\bs x},t) =  -K(\porosity) \nabla_{\bs{x}} p$, with the effective permeability $K$ defined in~\eqref{eq: def of K}. As for the permeability $K$, we note the right-hand side of~\eqref{eq: effective coefficient matrices} is a multiple of the identity matrix due to the symmetry of the cell problem~\eqref{eq: cell transport problem}.

\begin{figure}[t]
\centering
    	\includegraphics[width=\textwidth]{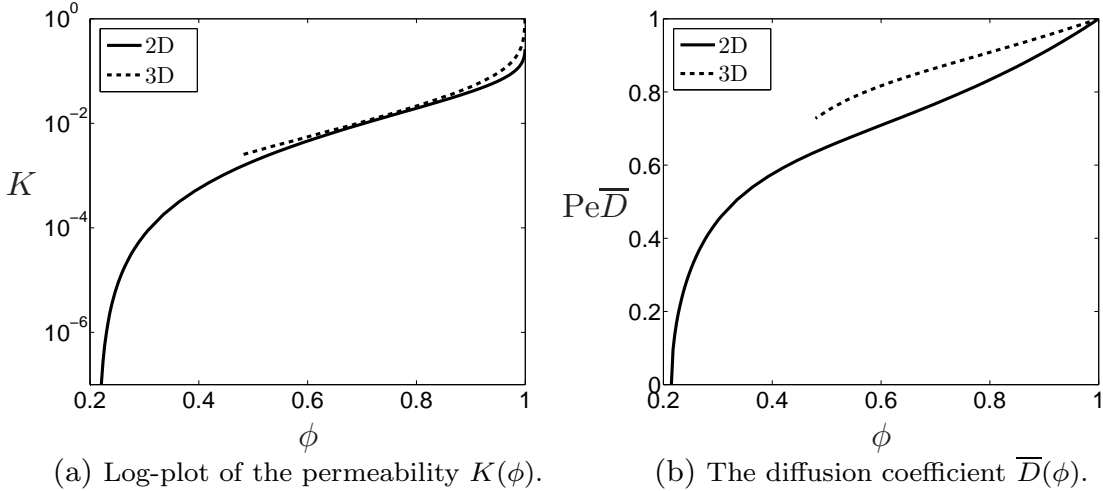}
\caption{The functions (a) $K(\porosity)$ and (b) $\Pec \Deff(\porosity)$ (defined in \eqref{eq: def of K} and \eqref{eq: effective coefficient matrices} respectively) calculated using Comsol Multiphysics, for ball obstacles whose centres are located in a cubic lattice, but whose radii can vary as a function of space. Due to the cubic lattice arrangement, the minimum porosity is $1 - \pi/4 \approx 0.21$ in two dimensions, and $1-\pi/6 \approx 0.48$ in three dimensions. Note that $K \to \infty$ as $\porosity \to 1$, but we bound the graph for illustrative purposes.}
\label{fig: Perm and Diff coeff}
\end{figure}

The effective permeability $K(\phi)$ and the effective diffusion $\Deff(\porosity)$ can be computed by solving the respective cell problems \eqref{eq: Flow problem  order 1 after scalings} and \eqref{eq: cell transport problem} for a given cell porosity $\phi$, determined by the size of the spherical obstacle. We do this numerically using the finite-element software Comsol Multiphysics and find that both $K$ and $\Deff$ increase as the porosity increases, as expected (see figure~\ref{fig: Perm and Diff coeff}). 
The additional degree of freedom afforded by the three-dimensional problem means that the permeability and effective diffusion are larger in three than in two dimensions. Our calculations show that the leading-order effective diffusion coefficient is not affected by the fluid flow, hence the effective diffusion shown in figure~\ref{fig: Perm and Diff coeff}(b) is equivalent to that calculated in \cite{Bruna2015diffusion}, which considered the limit of $\UU \to \zv$, $\kk \to 0$.

The homogenization procedure has significantly decreased the geometrical complexity of the multiply-connected domain in the full problem~\eqref{eq: transport problem ND}, while only slightly increasing the complexity of the governing equations for the homogenized problem~\eqref{eq: effective final problem}. This complexity manifests in the variable coefficients of the governing equation; these coefficients reflect the microscale structure of the problem in a systematic manner.

In addition to the increase in computational efficiency, our homogenized model gives insight into the physical effect of a non-uniform porosity. Namely, the homogenized governing equation~\eqref{eq: effective volumetric equation} indicates that a gradient in the porosity enters the equation as a term similar to the advection term due to the Darcy velocity. In other words, a non-uniform porosity in the membrane gives us an extra degree of freedom with which to either enhance or reduce the effect of a fluid flow on the solute concentration distribution. In the next section we determine how exactly to vary a macroscale porosity gradient to improve the effectiveness of a membrane filter.

\section{A uni-directionally graded filter}
\label{sec: Uni-dir problem}
\subsection{Model set-up}
\label{sec: Model set-up}

In this section we use the homogenized equations to understand and quantify the effect of porosity gradients on filter efficiency. We model a filtration process using~\eqref{eq: effective final problem} in the steady state, \ie setting $\partial \creal/\partial t =0$, coupled with appropriate boundary conditions at the inlet and outlet that represent two reservoirs. Since we are interested in the interaction between the fluid flow and the porosity gradient in the advection term, we consider a filter that is graded in the same direction as the fluid flow. Whilst it can be difficult to fine-tune filter porosity due to manufacturing constraints, it is acknowledged that a negative porosity gradient can improve filter efficiency. As discussed in the Introduction, a quantitative understanding of this behaviour is an open question.

Specifically, we consider an industrial process whereby a filter with a \emph{uni-directional} constant porosity gradient separates two reservoirs, and a flow is induced through the filter in the same direction as the gradient in porosity. We are interested in the steady-state operation of this process, and how the porosity gradient affects particle adsorption. We define the direction of porosity gradient as $x$, where $x \in (0, 1)$ within the filter (since the dimensional characteristic length $\leng$ is chosen to be the length of the filter). Thus the set-up is similar to that illustrated in figure~\ref{fig: Set-up schematic}. The upstream is defined for $x \in (-\infty,0)$ and the downstream for $x \in (1,\infty)$. We emphasize that the full problem occurs within a \emph{three-dimensional} domain (and take $d = 3$ henceforth) and it is only the variation in porosity that is one-dimensional.

Provided the boundary conditions allow for uni-directionality, the Darcy equations \eqref{eq: Darcy velocity} yield a uni-directional constant macroscale flux $\UU = U \bs{e}_x$ (where $\bs{e}_x$ is the unit vector in the $x$-direction), at all points in the filter, regardless of the porosity. We are able to take $U = 1$ without loss of generality by choosing the characteristic velocity scale $\velscale$ to be the average velocity in the $x$-direction, a quantity that is easily measured experimentally. The homogenized governing equation for the concentration $\creal$ within the filter \eqref{eq: effective volumetric equation} is then
\begin{align}
\label{eq: example governing equation}
\dbyd{}{x} \left[\Deff(\porosity) \dbyd{\creal}{x} - \dfrac{\creal}{\porosity}\left(1 + \Deff(\porosity) \dbyd{\porosity}{x} \right)\right] = f(\porosity) \creal, \qquad x \in (0,1).
\end{align}
Considering the limit of \eqref{eq: effective coefficient matrices} and \eqref{eq: example governing equation} as $\porosity \to 1$ provides governing equations for the upstream $\cus$ and downstream $\cds$ concentrations,
\begin{subequations}
\label{eq: additional systems}
\begin{alignat}{2}
\label{eq: additional systems a}
\dbyd{}{x} \left(\Pec^{-1} \dbyd{\cus}{x} - \cus \right) &= 0,& \qquad &x \in (-\infty,0), \\
\label{eq: additional systems b}
\dbyd{}{x} \left(\Pec^{-1} \dbyd{\cds}{x} - \cds\right) &= 0,& \qquad &x \in (1,\infty).
\end{alignat}
\end{subequations}

We impose continuity of concentration and concentration flux at the boundaries between the filter and the reservoirs. In the far-field of the reservoirs, the concentration tends to a constant value. We may take the upstream concentration $\cus \to 1$ (by choice of our nondimensionalization), while the downstream concentration $\cds$ tends to a constant value that must be determined as part of the solution. Mathematically, this corresponds to
\begin{subequations}
\label{eq: additional systems BC}
\begin{alignat}{2}
\label{eq: x to minus inf BC}
\cus &\to 1, &\qquad &x \to -\infty, \\
\label{eq: cont of c x=0}
\cus &= \dfrac{\creal}{\porosity}, &\qquad &x =  0, \\
\Pec^{-1} \dbyd{\cus}{x} - \cus &= \Deff(\porosity) \dbyd{\creal}{x} - \dfrac{\creal}{\porosity}\left(1 + \Deff(\porosity) \dbyd{\porosity}{x} \right), &\qquad &x =  0, \\
\label{eq: DS conc}
\cds &= \dfrac{\creal}{\porosity}, &\qquad &x =  1, \\
\label{eq: cont of flux x=L}
\Pec^{-1} \dbyd{\cds}{x} - \cds &= \Deff(\porosity) \dbyd{\creal}{x} - \dfrac{\creal}{\porosity}\left(1 + \Deff(\porosity) \dbyd{\porosity}{x} \right), &\qquad &x =  1, \\
\label{eq: x to inf BC}
\dbyd{\cds}{x} &\to 0, &\qquad & x \to \infty.
\end{alignat}
\end{subequations}

Equations~\eqref{eq: additional systems}, \eqref{eq: x to minus inf BC} and~\eqref{eq: x to inf BC} give
\begin{align}
\cus = 1 - A \exp(\Pec \, x), \quad \cds = B,
\label{eq: sol to up and downstream conc}
\end{align}
where $A, B \in [0, 1]$ are two constants to be determined. These expressions may be substituted into the remaining boundary conditions~\eqref{eq: cont of c x=0}--\eqref{eq: cont of flux x=L} to yield two boundary conditions for $\creal$, thus closing the elliptic problem~\eqref{eq: example governing equation},
\begin{subequations}
\label{eq: Correct BC}
\begin{alignat}{2}
\Deff(\porosity) \dbyd{\creal}{x} - \dfrac{\creal}{\porosity}\left(1 + \Deff(\porosity) \dbyd{\porosity}{x} \right) &= -1, &\qquad &x =  0, \\
\dbyd{\creal}{x} - \dfrac{\creal}{\porosity} \dbyd{\porosity}{x} &= 0, &\qquad &x =  1.
\end{alignat}
\end{subequations}
We solve the system for the filter~\eqref{eq: example governing equation} and \eqref{eq: Correct BC} numerically using a second-order accurate finite-difference scheme with $10^3$ grid points. To ensure that our results are experimentally relevant, we make comparisons between simulations that use the same transmembrane pressure difference. This corresponds to modifying the dimensionless parameters $\Pec$ and $\kk$ between simulations, and is discussed in more detail in Appendix~\ref{App: Parameter modification}.

\begin{figure}[t]
\centering
    	\includegraphics[width=\textwidth]{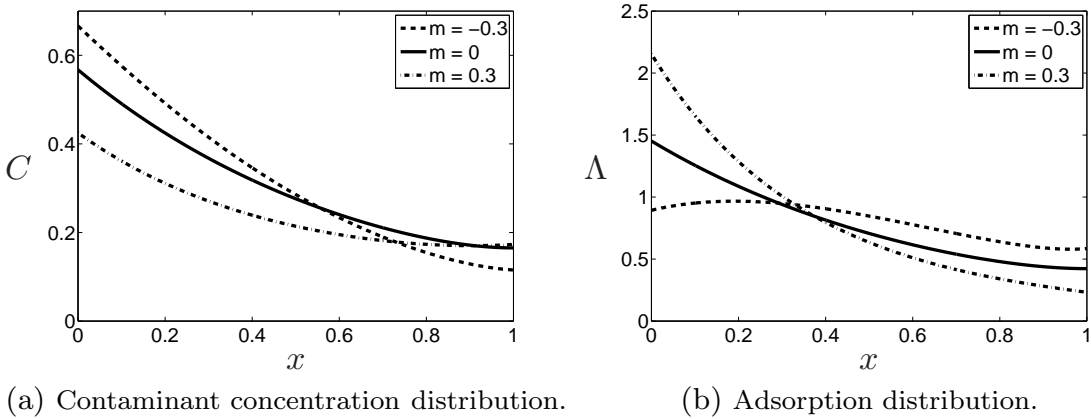}
\caption{Numerically determined concentration and particle removal profiles for the porosity function \eqref{porosity model} using $\porosity_0 = 0.75$ with three different gradients, $m=-0.3, 0, 0.3$. We use the reference values $\porosity_0 = 0.75$, $m = 0$, $\Pec = 3$, $\kk = 1$ from which to modify appropriate parameters (as discussed in Appendix~\ref{App: Parameter modification}). The concentration $\creal(x)$ is obtained from~\eqref{eq: example governing equation} and \eqref{eq: Correct BC} using a second-order accurate finite-difference scheme, from which the adsorption distribution $\uptake(x)$ can be deduced using~\eqref{eq: Uptake}.}
\label{fig: Volumetric_v_intrinsic}
\end{figure}

We consider a linear porosity model of the form 
\begin{equation}
\label{porosity model}
\porosity(x) = \porosity_0 + m(x - 0.5),
\end{equation}
for various mean porosities $\porosity_0$ and gradients $m$. This corresponds to a uniform filter when $m=0$. As we expect, the concentration $\creal$ decreases with $x$ as the contaminants are trapped by the filter medium (figure~\ref{fig: Volumetric_v_intrinsic}a). We also observe that the rate at which the concentration changes in $x$ (which is related to the rate of particle removal) decreases as the porosity gradient increases (figure~\ref{fig: Volumetric_v_intrinsic}a). We probe this result further in the next section, to explore which filter set-up is most efficient at contaminant removal.

\subsection{Filter efficiency}
\label{sec: Parameter sweep}

Since the goal of filtration is to maximize particle adsorption, it is helpful to consider the particle adsorption distribution, described by the right-hand side of \eqref{eq: example governing equation}, within the filter. To this end, we introduce 
\begin{align}
\label{eq: Uptake}
\uptake(x) = f\big(\porosity(x)\big) \creal(x).
\end{align}
This measures the concentration of contaminant removed per unit time at position $x$ in the filter. For the concentration profiles depicted in figure~\ref{fig: Volumetric_v_intrinsic}a, we show the corresponding $\uptake$ in figure~\ref{fig: Volumetric_v_intrinsic}b for comparison. A natural measure to evaluate filter efficiency is the total contaminant removal by the filter per unit filter area per unit time, given by
\begin{align}
\label{eq: T metric}
\total \equiv \int_0^1 \! \uptake(x) \, \mathrm{d}x. 
\end{align}
Using~\eqref{eq: example governing equation}, \eqref{eq: DS conc}, and \eqref{eq: Correct BC} we see that
\begin{align}
T = 1 - \cds \leq 1,
\end{align}
and thus total contaminant removal is directly related to outlet concentration, as expected.

\begin{figure}[t]
\centering
    	\includegraphics[width=\textwidth]{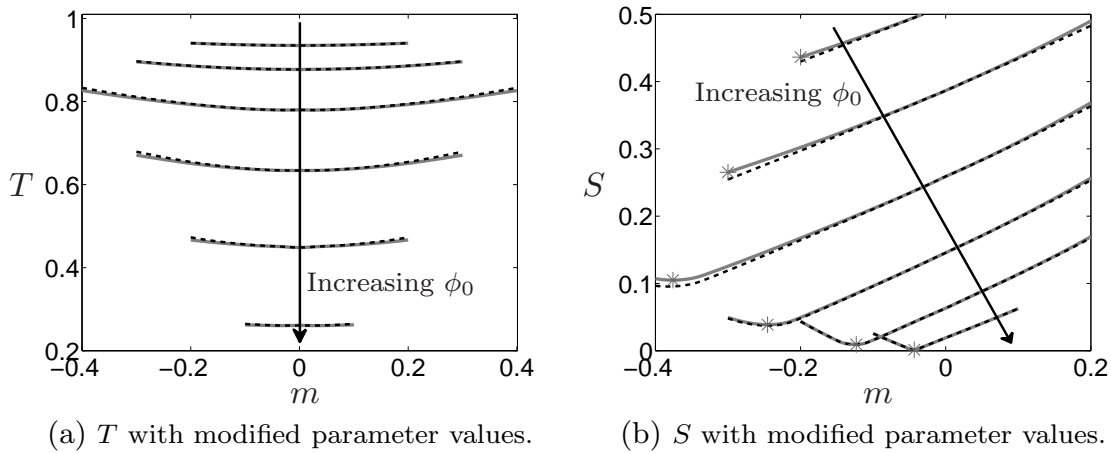}
\caption{Plots of the filter efficiency metrics $\total$ and $\mingrad$, defined in~\eqref{eq: T metric} and~\eqref{eq: M metric} respectively, for varying $\porosity_0$ and $m$ in $\porosity(x) = \porosity_0 + m(x - 0.5)$. Each grey curve represents a different value of $\porosity_0$, which increments in steps of $0.05$ from $\porosity_0 = 0.65$ to $\porosity_0 = 0.9$. We vary $m$ for a given $\porosity_0$ such that $\porosity \in [0.55,0.95]$ (which falls within the allowable range of $\porosity \in [1-\pi/6, 1]$ in three dimensions). Therefore, the available range of $m$ varies with $\porosity_0$. We use the reference values $\porosity = 0.75$, $\Pec = 3$, $\kk = 1$ from which to modify appropriate parameters (as discussed in Appendix~\ref{App: Parameter modification}). The asterisks in (b) denote the lowest value of $\mingrad$ for a varying $m$ with a fixed value of $\porosity_0$. The dashed black curves denote the asymptotic approximations of $\total$ and $\mingrad$ from results derived in \S\ref{sec: Asymptotic results}, and offer outstanding agreement with the numerical results at a fraction of the computational cost.}
\label{fig: M and T figures}
\end{figure}

We investigate how a porosity variation affects the total particle removal within a filter by calculating $\total$ as we vary the mean porosity $\porosity_0$ and gradient $m$ in the linear porosity function \eqref{porosity model}.  Although $\total$ is observed to increase with decreasing $\porosity_0$, as we would expect, there are diminishing returns as we continue to decrease $\porosity_0$ (see figure~\ref{fig: M and T figures}a). However, we observe only a weak dependence on $m$ (figure~\ref{fig: M and T figures}a) whereby, upon including a porosity gradient, there is an increase in the total particle adsorption $\total$. Interestingly, we find that the sign of $m$ does not affect $\total$, exhibited through the symmetry about $m = 0$. Whilst it may appear surprising that the orientation of the filter has no effect on total particle adsorption, this phenomenon arises as a result of the commutativity of the obstacle layout: the concentration after each subsequent obstacle is a proportion of the incoming concentration. Hence, the contaminant removal by the entire filter is a product of all these proportions and, moreover, any permutation of the sinks will result in the same total contaminant removal. If our model included pore blocking as a result of particle adsorption, this symmetry feature would no longer hold.

Thus, while the metric $\total$ is useful, it cannot account for the superior performance often observed for filters with a negative porosity gradient. To explain this experimental observation, we must investigate $\uptake$ further. Although long-term blockage is inevitable, a particular aim of an efficient filter is to avoid local blockages while the rest of the filter is still functional \ie to have the filter fail `all at once' rather than in one place. Thus, while the total contaminant removal is important, we must also consider the uniformity in which it is removed by the filter. To quantify this, we introduce the metric
\begin{align}
\label{eq: M metric}
\mingrad &= \int_0^1 \! \left|\uptake(x) - \total \right| \, \mathrm{d}x;
\end{align}
$\mingrad$ is minimized (and zero) for uniform contaminant removal and so we posit that a smaller value of $\mingrad$ corresponds to a more efficient filter.

In contrast to our analysis of the total particle adsorption, $\total$, we find that $\mingrad$ strongly depends on the orientation of the filter (see figure~\ref{fig: M and T figures}b). Additionally, a negative porosity gradient $m$ yields a concentration profile that is closer to providing spatially uniform contaminant removal than a positive $m$. We note however that the minimum value of $\mingrad$ for a given mean porosity $\porosity_0$ (shown as asterisks in figure~\ref{fig: M and T figures}b) does not necessarily correspond to choosing the porosity gradient as negative as possible. This feature occurs because a porosity gradient that is too negative will have less particle adsorption near the entrance to the filter than towards the exit, thus also creating an uneven adsorption distribution.

Our analysis therefore indicates that, for a given mean porosity, the porosity gradient has a much larger effect on the distribution of particle adsorption within the filter than it does on the total particle adsorption. This provides a quantitative interpretation for the experimental observation that efficiency in contaminant removal is improved when a filter with a negative porosity gradient is used. As the shape of the concentration profile relates to the problem of localized blocking, we hypothesize that a negative porosity gradient allows for a more uniform contaminant removal. This increases the time until localized blocking becomes a problem, and thus increases filter efficiency.

\subsection{Slowly varying porosity}
\label{sec: Asymptotic results}

The results we have presented so far provide a very useful quantitative analysis on the role of porosity gradients in filtration. However, because they were obtained numerically, it is still rather cumbersome to perform major parameter sweeps through the space of porosity functions and the dimensionless parameters $\Pec$ and $\kk$ (although we emphasize that results we presented in the previous section are still orders of magnitude cheaper than solving the full equations in a complex geometry).

In this section, we take a further simplifying step by exploiting the fact that manufacturing constraints limit the possible variations in porosity. We determine approximate analytic solutions to the equations presented, significantly increasing the computational efficiency and allowing us to gain insight into the concentration profiles we obtain. To implement this, we explore the limit of a weakly varying porosity such that $\mathrm{d}\porosity/\mathrm{d}x = \order{\epsb}$, where $\epsb \ll 1$. 

We expand $\creal$ and $\porosity(x)$ in powers of $\epsb$ as follows
\begin{align}
\label{eq: asym series weak porosity}
\creal(x) \sim \creal_0(x) + \epsb \creal_1(x), \quad \porosity(x) \sim \porosity_0 + \epsb \porosity_1(x) \quad \text{as } \epsb \to 0,
\end{align}
where $\porosity_0$ is the mean porosity, thus constant in $x$. We expand $\Deff(\porosity(x))$ and $f(\porosity(x))$ in the same manner as $\porosity(x)$, \ie where the leading-order term is constant. The coefficients are given by
\begin{align}
\begin{aligned}
\Deff_0 &= \Deff(\porosity_0),& \quad \Deff_1(x) &= \porosity_1(x) \dbyd{\Deff}{\porosity}(\porosity_0), \\  f_0 &= \kk\dfrac{ d (1 - \porosity_0)}{\porosity_0}\left(\dfrac{V_d}{1- \porosity_0}\right)^{1/d},& \quad f_1(x) &= \kk \porosity_1(x) \dfrac{\porosity_0 - d}{\porosity^2_0}\left(\dfrac{V_d}{1- \porosity_0}\right)^{1/d}.
\end{aligned}
\end{align}
As the porosity function is an input to our model, the asymptotic expansions for $\porosity(x)$, $\Deff(\porosity(x))$, and $f(\porosity(x))$ are known. For the linear porosity functions~\eqref{porosity model} considered in the previous section, we have $\epsb = m$, $\porosity_1(x) = x - 0.5$, and the asymptotic limit we consider is equivalent to $m \to 0$, \ie close to uniform porosity. Note that the theoretical maximum value of $\epsb$ for a linear porosity function is $\pi/6 \approx 0.52$ in three dimensions. However, the maximum value of $\epsb$ will be smaller in practise.

Substituting the asymptotic series~\eqref{eq: asym series weak porosity} into the system \eqref{eq: example governing equation}, \eqref{eq: Correct BC} and equating powers of $\epsb$, the $\order{1}$ system is given by
\begin{subequations}
\label{eq: order 1 weak porosity}
\begin{alignat}{2}
\dbyd{}{x}\left (\Deff_0 \dbyd{\creal_0}{x} - \dfrac{\creal_0}{\porosity_0} \right) &= f_0 \creal_0, &\qquad &x \in (0,1), \\
\Deff_0 \dbyd{\creal_0}{x} - \dfrac{\creal_0}{\porosity_0} &= -1, &\qquad &x =  0, \\
\dbyd{\creal_0}{x} &= 0, &\qquad &x =  1.
\end{alignat}
\end{subequations}
The system~\eqref{eq: order 1 weak porosity} is solved by
\begin{subequations}
\label{eq: C_0 asy}
\begin{align}
\creal_0 = 2 \alpha \porosity_0 \exp(a x)\left[b \cosh ab(x-1) - \sinh ab(x-1)\right],
\end{align}
where
\begin{align}
a = \dfrac{1}{2 \porosity_0 \Deff_0}, \quad
b = \sqrt{1 + f_0 /\left(a^2 \Deff_0 \right)}, \quad
\alpha = \dfrac{1}{(1+b^2)\sinh ab + 2 b \cosh ab}.
\end{align}
\end{subequations}

The $\order{\epsb}$ terms in the system \eqref{eq: example governing equation} and \eqref{eq: Correct BC} are given by
\begin{subequations}
\label{eq: order eps weak porosity}
\begin{alignat}{2}
\dbyd{}{x} \!\left (\Deff_0 \dbyd{\creal_1}{x} - \dfrac{\creal_1}{\porosity_0}  +  \Deff_1(x) \dbyd{\creal_0}{x} - \dfrac{\creal_0}{\porosity_0}\! \left(  \Deff_0 \dbyd{\porosity_1}{x} - \dfrac{\porosity_1}{\porosity_0}  \right)  \! \right)  
  &= f_0 C_1+ f_1(x) \creal_0 , & &x \in (0,1), \\
\Deff_0 \dbyd{\creal_1}{x} - \dfrac{\creal_1}{\porosity_0} + \Deff_1(0) \dbyd{\creal_0}{x} - \dfrac{\creal_0}{\porosity_0}\! \left(  \Deff_0 \dbyd{\porosity_1}{x} - \dfrac{\porosity_1}{\porosity_0}  \right) & =0 , &\  &x =  0, \\
\dbyd{\creal_1}{x} &= \dfrac{\creal_0}{\porosity_0} \dbyd{\porosity_1}{x}, &\quad &x =  1.
\end{alignat}
\end{subequations}
Using variation of parameters, the system~\eqref{eq: order eps weak porosity} is solved by
\begin{align}
\label{eq: creal_1 asy}
\creal_1(x) &= \dfrac{\exp(ax)}{\sinh ab}\bigg[\left(A_1(x) + \alpha B_1\right) \sinh abx + \left(A_2(x) + \alpha B_2\right) \sinh ab\left(x -1\right)\bigg].
\end{align}
Here, 
\begin{subequations}
\label{eq: Asymptotic coefficients}
\begin{align}
A_1(x) &= \dfrac{1}{a b \Deff_0}\int_x^1 \! g(t,\creal_0(t)) \exp(-at) \sinh ab(t-1)\, \mathrm{d}t , \\
A_2(x) &= \dfrac{1}{a b \Deff_0}\int_0^x \! g(t,\creal_0(t)) \exp(-at) \sinh abt \, \mathrm{d}t, \\
\begin{pmatrix}
  B_1 \\
  B_2
 \end{pmatrix}
 &=
 \begin{pmatrix}
  -q & b \\
  b & -q
 \end{pmatrix}
 \begin{pmatrix}
  b \left(A_2(1) - 2 \dbyd{\porosity_1(1)}{x}\right) \\
  b A_1(0) + N(0,\creal_0(0))/(a \Deff_0)
 \end{pmatrix}, \\
 g(x,\creal(x)) &= -\dbyd{}{x}N(x,\creal(x)) +f_1(x) \creal(x), \\
 N(x,\creal(x)) &= \Deff_1(x)\dbyd{\creal}{x} - \dfrac{\creal(x)}{\porosity_0}\left(\Deff_0 \dbyd{\porosity_1}{x} - \dfrac{\porosity_1(x)}{\porosity_0} \right), \\
 q &= \sinh ab + b \cosh ab.
\end{align}
\end{subequations}

\begin{figure}[t]
\centering
    	\includegraphics[width=\textwidth]{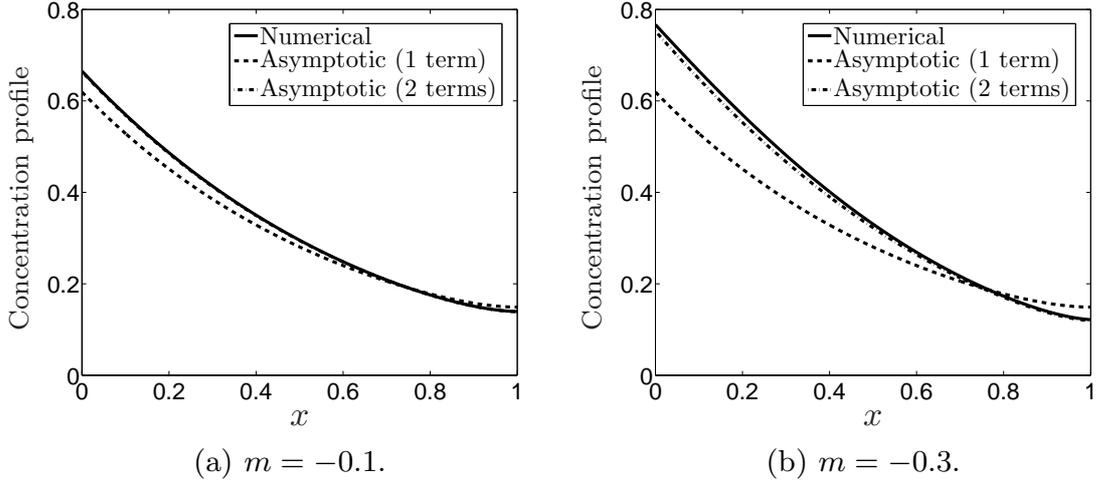}
\caption{The volumetric concentration profiles generated from numerical (solving the system~\eqref{eq: example governing equation}, \eqref{eq: Correct BC} using a finite-difference scheme) and asymptotic (obtained from~\eqref{eq: C_0 asy} and~\eqref{eq: creal_1 asy}) methods. The parameter values used are $\Pec = 5$, $\kk = 1$, $\porosity(x) = 0.75 + m(x - 0.5)$. The two-term asymptotic solution is indistinguishable from the numerical solution in (a).}
\label{fig: Asy v num}
\end{figure}

In addition to the increase in computational efficiency, this analytic result also allows us to determine closed-form approximations up to $\order{\epsb}$ for the metrics $\total$ and $\mingrad$,
\begin{subequations}
\label{eq: closed form metrics}
\begin{align}
\total &\sim 1 - \dfrac{\alpha \exp(a)}{\porosity_0} \left(2 b \porosity_0 + \epsb \left(  \dfrac{B_1}{\sinh ab} - 2 b \porosity_1(1) \right)\right), \\
\label{eq: closed form mingrad}
\mingrad &\sim  \int_0^1 \! \left| f_0 \creal_0(x) + \epsb \left(f_1(x) \creal_0(x) + f_0 \creal_1(x) \right) - \total\right|\, \mathrm{d}x,
\end{align}
\end{subequations}

The asymptotic concentration profile (up to two terms) agrees very well with the profile obtained numerically, even when the porosity gradient takes physically extreme values (see figure~\ref{fig: Asy v num}). Further, the closed-form nature of~\eqref{eq: closed form metrics} allows us to produce approximate results for $\total$ and $\mingrad$ (shown as dashed black lines in figure~\ref{fig: M and T figures}) at a fraction of the computational cost. As with all the analytic results in this section, we observe excellent agreement with the numerical results.

\section{Discussion}
\label{sec: Discussion}

We have systematically derived a macroscopic model for a porosity-graded filter from microscale information by generalizing standard homogenization theory for near-periodic systems. The result is an advection--diffusion--reaction equation for the solute concentration within the filter as a function of the porosity distribution and operating conditions. In particular, this equation has allowed us to investigate how porosity gradients affect solute trapping and filter efficiency. To this end, we have defined two suitable metrics to quantify the implicit effect of blocking. The first corresponds to the rate of total contaminant removal. The second  measures how close to uniform the adsorption of the contaminant is within the filter and is an indication of the propensity of a filter to localized blocking. 
By performing a computationally efficient parameter sweep, we have been able to determine and quantify that, for a given mean porosity, a porosity gradient has a much larger effect on the second metric than the first metric. In general, we have found that a filter with a decreasing porosity will be less prone to localized blocking and we have quantified the optimal gradient as a function of the operating conditions. This allows us to understand the experimental observation that a decreasing porosity can lead to a greater filter efficiency; the porosity gradient lowers the risk of localized blocking, while maintaining the rate of total contaminant removal. 

The homogenization procedure we used allowed a near-periodic, rather than a strictly periodic, microstructure to be considered. The macroscale variation that near-periodicity allows is a vital feature of a porosity-graded filter. The resulting macroscale equations are computationally cheap to solve, allowing us to efficiently explore the experimental parameter space to quantify filter efficiency.

With regards to an `ideal' filter, we note that it is impossible to globally optimize both of the two metrics ($\total$ and $\mingrad$) that we created. Indeed, we found that porosity functions that increase total adsorption were further from uniform adsorption. Thus, there is a trade-off to be made between maximizing adsorption and minimizing potential blockage issues, an idea supported in the network model for trapping~\citep{griffiths2014combined}. Whilst the question of determining an `optimum' porosity function will depend on the specifications of the end-user, our work provides an inexpensive method to increase filter efficiency given these requirements. However, our model does suggest that, in the absence of additional constraints, it is wiser to use a filter with a negative rather than positive porosity gradient. While this will not affect the total adsorption, it will provide a more uniform adsorption rate throughout the filter, which should reduce blocking issues.

This work can also be used to investigate filters comprised of a series of material layers, each with a different mean porosity. One way to achieve this would be to derive suitable boundary conditions to couple each of the regions, using the method in \S\ref{sec: Model set-up}. Another (simpler) idea, would be to use a differentiable approximation of the porosity (\eg cubic splines defined between the centres of neighbouring layers) in the macroscale model defined in the first part of this paper.


Although we considered a filter whose microscale structure consisted of balls, it is possible to consider an arbitrary shape for the microscale obstacle. However, we chose our particular microstructure to allow an explicit macroscale equation, and thus significant analytic progress to be made, and the same will not hold for the general microstructure case. It is a simple task to combine this work with that of~\citep{richardson2011derivation}, who considered a general curvilinear coordinate transform to map a near-periodic microscale to a periodic domain, thus allowing a homogenization method to be applied. However, as the coefficients within the resulting governing equations would have to be determined at each point in space, the `reduced' equations are still computationally expensive.

One drawback of the method presented here is the restriction to a near-periodic microscale, meaning that materials that vary wildly on the microscale cannot be treated similarly. A viable extension to this work would be to test how robust the assumptions on a near-periodic microscale structure are, and how far away one could get before the macroscopic description we derived broke down. Although the near-periodic assumption is required to derive the macroscale equations, it may be the case that it is only the average properties of the microstructure that must possess this property. This could be tested by performing numerical simulations on the full problem for randomly placed spheres. Similar simulations have been carried out in~\citep{Bruna2015diffusion} for diffusing particles only, and it was found that the concentration distribution for a random and cubic array of spheres agree in the high-porosity limit. On physical grounds, we expect the same result for our work in the high-porosity limit. Furthermore, in the same limit, we expect the results for a general microscale obstacle to coincide with the high-porosity limit of the governing equations derived in this paper, using appropriate results for pore volumes and obstacle surface areas.

One feature that is lacking in our model is the dynamic effect of blockage (membrane fouling). As blocking would temporally change the geometry of the problem, the flow and particle transport problems would be fully coupled, and hence difficult to investigate. Despite the lack of explicit blocking in our model, we were able to implicitly infer the long-term effect this would have by quantifying the particle adsorption distribution within the filter. If blocking was also considered, we would expect the results in this paper to provide a quasi-static description of that model. Additionally, the linear trapping rate that we imposed is a generalized approximation of a feature that will depend on the solute solubility and the filter material adsorption. Intuitively, one would expect a Michaelis--Menten type dependence for the adsorption rate on the number of particles, \ie approximately linear for a small number of particles and bounded above for a large number of particles. Our approach is valid because of the dilute suspension assumption but, if one is in the same asymptotic limit, it is a trivial task to extend the analysis in this paper to a general nonlinear adsorption rate if required.

Finally, we note that this work has the potential not only to guide filter manufacture and operating conditions, but also to provide assistance to other industries that use functionally graded materials, such as heterogeneous artificial body tissue in tissue engineering~\citep{weiss2005bayesian,sun2004computer}, or graded electrodes in lithium-ion batteries~\citep{chung2014particle}. Furthermore, as the technology to produce functionally graded materials grows~\citep{kieback2003processing}, the potential experimental parameter space will also increase. Forming appropriate mathematical models and maximizing the analytic progress that can be made will significantly expedite the exploration of this parameter space and result in faster technological growth.
\vskip6pt

\appendix

  \renewcommand{\theequation}{A\arabic{equation}}
  \setcounter{equation}{0}  
  
\section{Transport theorem for a varying domain}
\label{App: TT}

During the homogenization procedure, we use the transport theorem to evaluate integrals over microscale cells which vary in the macroscale variable $\bs{x}$. 
To do this, we must determine the rate of change of the fluid domain $ \regf(\bs{x})$ with respect to $\bs{x}$.

We note that $ \regf(\bs{x})$ has a fixed outer boundary, $\partial \rege$, and an interior boundary on the surface of the obstacle of radius $R({\bs x})$, $\partial \regs (\bs{x})$. The rate of change of $\partial \regs (\bs{x})$ with respect to $\bs{x}$ is $\nabla_{\bs{x}} \eps$. To see this, consider the difference between integrals over the domains $\regf(\bs{x} + \xi \bs{e}_i)$ and $\regf(\bs{x})$ as $\xi \to 0$. The resulting domain of integration is a shell whose thickness is approximately $\xi \partial \eps / \partial x_i$ as $\xi \to 0$. As we are considering an integral over $\regf(\bs{x}) = \rege(\bs{x}) \backslash \regs(\bs{x})$, the relevant velocity of the interior boundary is $-\nabla_{\bs{x}} \eps$.

Therefore, the transport theorem for the geometry considered in this paper is as follows. Let $\bs{v}(\bs{x},\bs{y},t)$ be a continuous vector field and periodic in $\bs{y}$ on the outer boundary $\partial \omega(\bs{x})$. Then
\begin{align}
\label{eq: TT}
\nabla_{\bs{x}} \bcdot \int_{\regf (\bs{x})} \! \bs{v}(\bs{x},\bs{y},t) \, \mathrm{d}\bs{y} = \int_{\regf(\bs{x})} \! \nabla_{\bs{x}} \bcdot \bs{v}(\bs{x},\bs{y},t) \, \mathrm{d}\bs{y}
 - \int_{\partial \regs(\bs{x})} \! \nabla_{\bs{x}} \eps \bcdot \bs{v} (\bs{x},\bs{y},t) \, \mathrm{d}s.
\end{align}

  \renewcommand{\theequation}{B\arabic{equation}}
  \setcounter{equation}{0}  

\section{Parameter modification}
\label{App: Parameter modification}

Whilst the two dimensionless parameters $\Pec$ and $\kk$ are useful for a mathematical analysis of the governing equations, within an experimental set-up it is not these parameters that will, in general, be kept constant as we vary the membrane porosity. When analyzing results from our model, it is useful to consider mathematical variations that correspond to experimental variations in order to maximize the experimental relevance of our work.

In practice, the control parameter within a filtration system is typically the transmembrane pressure difference (note that the same calculation applies if the volumetric flux is the control parameter that is used instead). This corresponds to a linear variation in the average velocity in the $x$-direction, $\velscale$. For a given (dimensional) transmembrane pressure difference $\pdiff$, we can solve~(\ref{eq: Darcy velocity}a) and (\ref{eq: Darcy velocity}c) in one dimension to deduce that (in dimensional variables)
\begin{align}
\label{eq: Velocity scale}
\velscale[\porosity(x)] = \dfrac{(\epsa \leng)^2 \pdiff}{\viscosity}\left(\displaystyle\int_0^\leng \! \dfrac{\mathrm{d}x}{K(\porosity(x))}\right)^{-1}.
\end{align}
Here, the square brackets denote that $\velscale$ is a functional of $\porosity(x)$.

\begin{figure}[t]
\centering
    	\includegraphics[width=\textwidth]{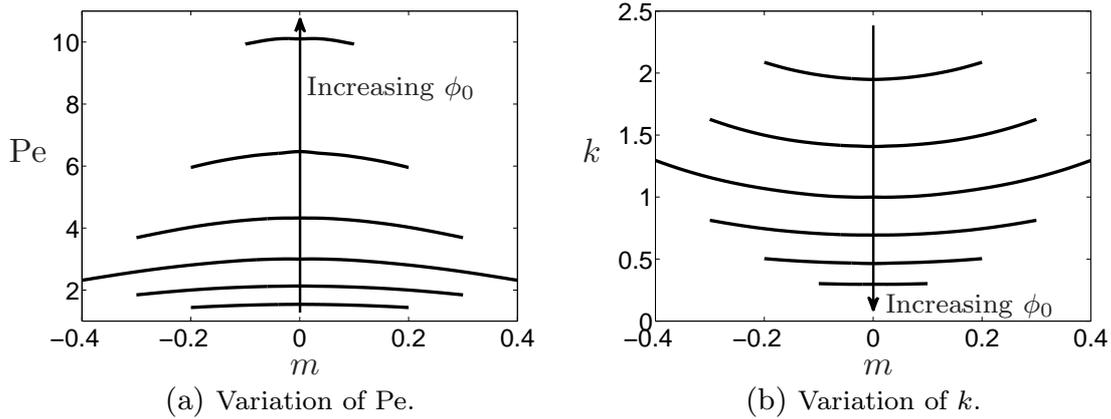}
\caption{Plots of how the dimensionless parameters $\Pec$ and $\kk$ vary in a manner consistent with keeping the transmembrane pressure difference constant between experiments for varying $\porosity_0$ and $m$ in $\porosity(x) = \porosity_0 + m(x - 0.5)$. Each curve represents a different value of $\porosity_0$, which increments in steps of $0.05$ from $\porosity_0 = 0.65$ to $\porosity_0 = 0.9$. We vary $m$ for a given $\porosity_0$ such that $\porosity \in [0.55,0.95]$ (which falls within the allowable range of $\porosity \in [1-\pi/6, 1]$ in three dimensions). Therefore, the available range of $m$ varies with $\porosity_0$. We use the reference values $\porosity = 0.75$, $\Pec = 3$, $\kk = 1$. We use equation~\eqref{eq: Velocity scale} to determine the value of $\velscale$ for a given porosity function, then use this to calculate the appropriate values of $\Pec$ and $\kk$, as discussed in the main text.}
\label{fig: Varying Pe and k figures}
\end{figure}

Therefore, we are able to determine how $\velscale[\porosity]$ varies (relative to a reference value for the transmembrane pressure difference) for different porosity functions $\porosity(x)$. That is, for a given reference function $\porosity_{\text{ref}}(x)$, and parameter values $\Pec[\porosity_{\text{ref}}]$ and $\kk[\porosity_{\text{ref}}]$, we can determine the effect of changing $\porosity(x)$ on $\velscale[\porosity]$ whilst keeping the transmembrane pressure difference constant. This, in turn, allows us to calculate the values of $\Pec[\porosity]$ and $\kk[\porosity]$ in the dimensionless model that correspond to the new function $\porosity$ while keeping the transmembrane pressure difference constant. Accordingly, when we discuss $\mingrad$ and $\total$ in \S\ref{sec: Parameter sweep}, we take $\Pec[\porosity]$ and $\kk[\porosity]$ to vary appropriately, relative to a reference point, defined by $\porosity_0 = 0.75$, $m=0$, $\Pec = 3$, $\kk = 1$ (illustrated in figure~\ref{fig: Varying Pe and k figures}).

\bibliographystyle{unsrtnat}
\small
\bibliography{ref.bib}

\end{document}